\documentclass[prl,twocolumn,showpacs,superscriptaddress]{revtex4}
\usepackage{graphicx,bm,amssymb,amsmath,slashed}

\newcommand{\Eq}[1]{Eq.~\eqref{#1}}
\newcommand{\eq}[1]{\eqref{#1}}

\newcommand{\beq}{\begin{equation}}
\newcommand{\eeq}{\end{equation}}

\newcommand{\beqa}{\begin{eqnarray}}
\newcommand{\eeqa}{\end{eqnarray}}

\newcommand{\Beqa}{\begin{eqnarray*}}
\newcommand{\Eeqa}{\end{eqnarray*}}

\newcommand{\nn}{\nonumber}

\newcommand{\vect}[1]{\mathbf{#1}}

\begin{document}
\newcommand{\cI}{{\mathcal I}}
\title{Universal Bound States of Two Particles in Mixed Dimensions or Near a Mirror}

\author{Shina Tan}
\affiliation{School of Physics, Georgia Institute of Technology, Atlanta, Georgia 30332, USA}
\affiliation{Center for Cold Atom Physics, Chinese Academy of Sciences, Wuhan 430071, China}

\begin{abstract}
Some novel \emph{two}-body effects analogous to the well-known \emph{three}-body Efimov effect
are predicted. In the systems considered,
particle A is constrained on a \emph{truncated} or \emph{bent} one-dimensional line or two-dimensional plane,
or on one side of a flat mirror in three dimensions (3D).
The constraining potential is fine-tuned such that particle A's ground state wave function is a constant
in the region in which it is constrained.
Particle B moves in 3D and interacts with particle A resonantly.
An infinite sequence of giant two-body bound states are found in each case.
%For cold atoms, such bound states will be free from three-body recombination if isolated from each other,
%and can potentially be very long lived.
\end{abstract}
\maketitle

%In the classical world, particles separated by distances beyond the range of forces between them
%can not be bound together. In quantum mechanics, two particles with a short interaction range $r_0$ and a large scattering length $a$ ($\gg r_0$)
%can, however, form a very shallow bound state in which the average distance between the two particles is of the order of
%$a$, if $a>0$. When $1/a=0$, such a two body bound state disappears, but three particles,
%such as three identical bosons as Efimov
%considered in 1970, can form a sequence of similar shallow bound states whose sizes are like $R\lambda^n$ at $1/a=0$. Here $n=$ integers,
%and $\lambda$ is a universal factor independent of the details of interactions at short distances. For indentical bosons, Efimov found $\lambda=22.7$.
%Efimov's effect has broad impact on nuclear theory, molecular halos, and ultracold atoms.
%It was recently confirmed directly in experiments on ultracold alkali atoms near Feshbach resonances.
Three particles with short range interactions and large scattering length usually
have many shallow bound states known as the Efimov states \cite{Efimov1970, Efimov1971,
EfimovExperiment2006, Braaten2007, EfimovExperiment20080603, EfimovExperiment20080722,
EfimovExperiment20081017, EfimovExperiment20090120, EfimovExperiment20090127, EfimovExperiment20090625, EfimovExperiment20091005,
EfimovExperiment20100325, EfimovExperiment20101010}.
Their sizes greatly exceed the range of the interactions, and
they are similar to each other due to a discrete scaling symmetry
 \cite{Efimov1970, Efimov1971,
EfimovExperiment2006, Braaten2007, EfimovExperiment20080603, EfimovExperiment20080722,
EfimovExperiment20081017, EfimovExperiment20090120, EfimovExperiment20090127, EfimovExperiment20090625, EfimovExperiment20091005,
EfimovExperiment20100325, EfimovExperiment20101010}.
Initially studied for three particles in three dimensions (3D),
Efimov's scenario has been extended to four or more particles in 3D \cite{cluster20060719,tetramer20061030, dimerdimer20080328,
tetramer20081029, tetramer20090306, tetramer20090319, dimerdimer20090406, EfimovExperiment20091005, cluster20100326,
tetramer20100624, cluster20110630, tetramer20110907},
to three or four resonantly interacting particles in mixed dimensions - where
different particles live in different spatial dimensions \cite{MixedD2008, MixedD20090321, MixedD20100126, MixedD20110413},
to five-body systems in one dimension (1D)\cite{Efimov1D2009}, and to three particles with long-range dipole interactions \cite{Dipole20110307}. 

All the above effects require \emph{three} or more interacting particles.
Although one may have Efimov effect for two particles interacting resonantly with a static point impurity
and between themselves,
the impurity can be treated as a third ``particle", having an infinite mass.

In this Letter we propose some novel scenarios in which one can \emph{not} say a third ``particle" (in the normal sense of the word) is involved,
and two particles only with \emph{short-range interactions} \cite{interaction} have many shallow bound states.
These two-body states are precisely similar to each other
and are universal in the sense that they depend only on such macroscopic parameters
as mass ratio and angle, and a single length scale that fixes their overall sizes, but not on any additional microscopic details of the problem.

\begin{figure}
\includegraphics[scale=0.29]{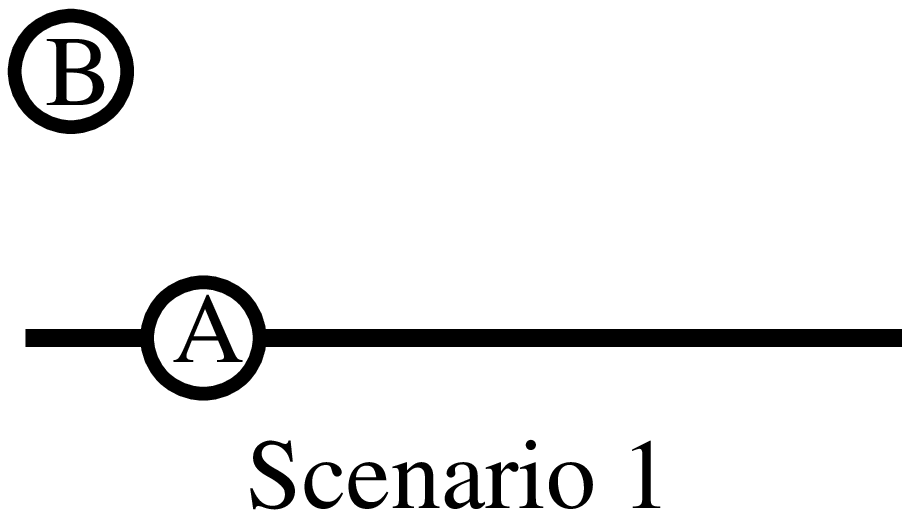}~\includegraphics[scale=0.29]{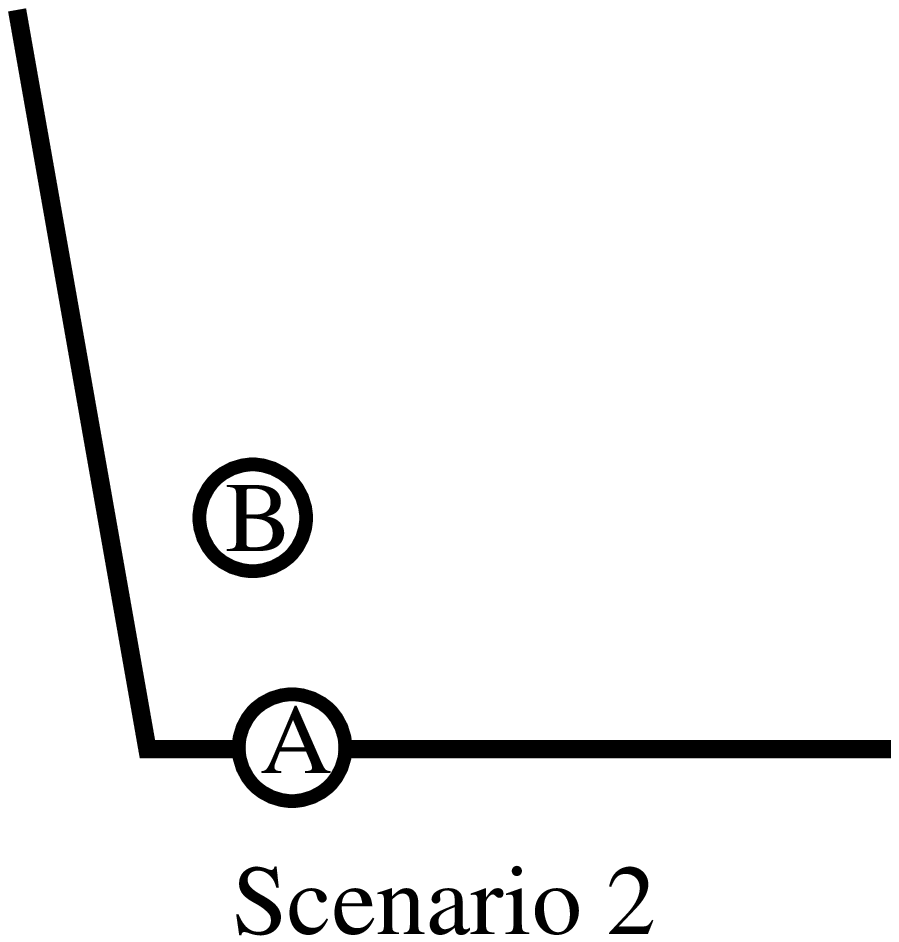}~~~~\includegraphics[scale=0.29]{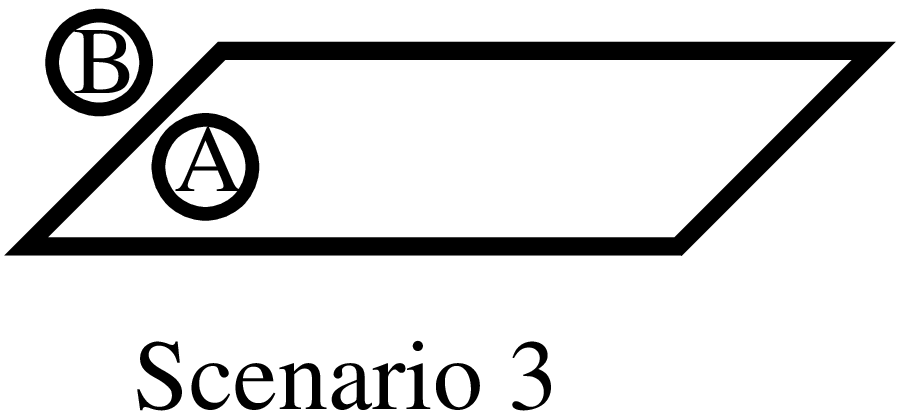}

\vspace{3mm}

\includegraphics[scale=0.29]{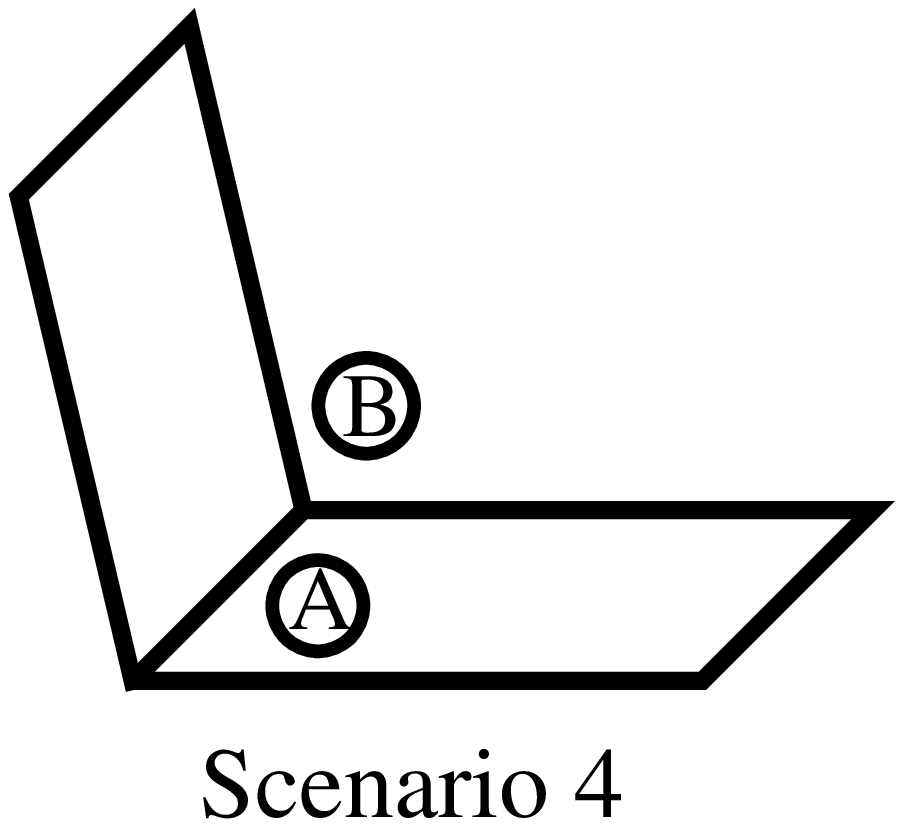}~~~~~~\includegraphics[scale=0.29]{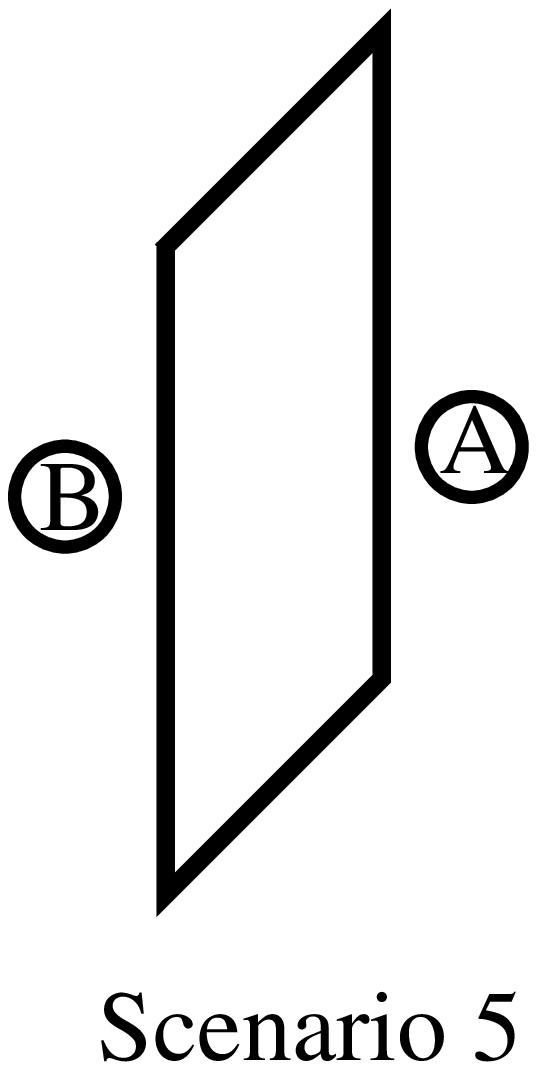}~~~~~~~~~~
\includegraphics[scale=0.29]{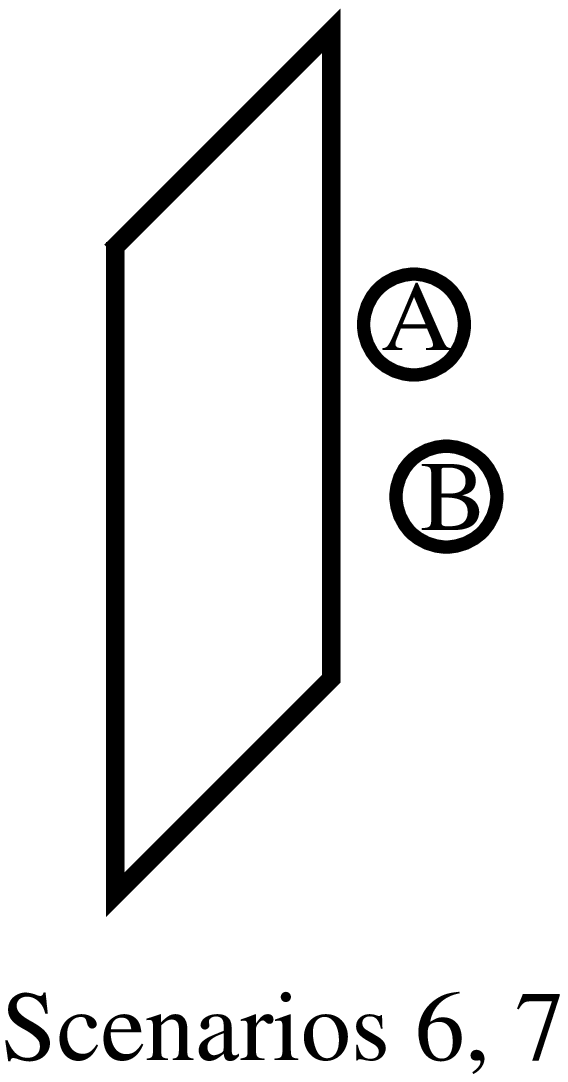}~~
\caption{\label{fig:sketch}
In Scenarios 1 and 2, particle A is constrained along a 1D ray or bent line. In Scenarios 3 and 4, particle A is constrained on a half-infinite plane or bent plane.
In Scenarios 5-7, particle A is constrained to one side of a plane in the 3D space.
In Scenarios 1-5, particle B is free to move in the entire 3D space.
In Scenarios 6 and 7, particle B is constrained to the same side of the plane as particle A. See text for more details.
}
\end{figure}

We propose 7 distinct but related scenarios (skteched in Fig.~\ref{fig:sketch}) for two particles A and B, with masses $m_A$ and $m_B$ respectively.

Scenario 1: A is constrained along a half-infinite 1D ray. The ray extends from a fixed point called \emph{vertex}.

Scenario 2: A is constrained along a V-shaped bent line, whose two straight arms form an angle $\theta<\pi$ at a point called \emph{vertex}.

Scenario 3: A is constrained on a half-infinite 2D planar sheet, whose \emph{edge} is a straight line.

Scenario 4: A is constrained on a bent 2D plane, whose two flat sheets form an angle $\theta<\pi$ along a 1D ridge called \emph{edge}.

Scenario 5: A is constrained to one side of a 2D plane (called \emph{mirror}) in 3D space.

Scenarios 6 and 7: A and B are \emph{both} constrained to the same side of a 2D plane (called \emph{mirror}) in 3D space.

In each scenario, the microscopic profile of the constraining potential is not uniquely specified.
We only require that the potential be fine-tuned and, in particular,
the \emph{depth} of the potential at the vertex, edge, or mirror surface must take some critical value (depending on the microscopic
size or thickness of the vertex, edge, or mirror surface), such that the ground state wave function
of an \emph{isolated} particle A approaches a finite \emph{constant} away from the vertex, edge, or mirror surface.

In Scenarios 1-5, B does not experience any constraining potential, and can move in the entire 3D space freely.
In Scenario 6, the constraining potential for particle B is also fine-tuned such that the ground state wave function
of an isolated particle B is a constant on one side of the mirror and zero on the other side.
In Scenario 7, the mirror acts as a hard wall for particle B, so that the ground state wave function of an isolated particle B
is linearly proportional to its distance from the mirror.

%In Fig.~\ref{fig:sketch} all the scenarios are sketched.

In Scenarios 1 and 2, A and B interact with a 1D-3D mixed dimensional scattering length $a_\text{eff}$ \cite{MixedD2008}.
In Scenarios 3 and 4, A and B interact with a 2D-3D mixed dimensional scattering length $a_\text{eff}$ \cite{MixedD2008}.
In Scenarios 5-7, A and B interact with a scattering length $a$.

In this Letter we show that, if $a_\text{eff}$ (for Scenarios 1-4)
 or $a$ (for Scenarios 5-7) is infinite, the two particles 
have an infinite number of shallow bound states that are similar to each other,
and we determine their wave functions \emph{analytically}.
The linear sizes of these bound states form a geometric sequence with common ratio
\beq\label{lambda}
\lambda=e^{\pi/s_0}
\eeq
called \emph{scaling factor} (greater than $1$),
and the binding energies form a geometric sequence with common ratio $\lambda^{-2}$, analogous to the Efimov effect \cite{Efimov1970, Efimov1971}.
Here $s_0$ is the positive solution to the transcendental equation
%\beq
%\frac{\sinh(s_0\alpha)}{s_0\sin\alpha}-\cosh(\pi s_0)+\eta(1)\Big\{1+\frac{\sinh[(\pi-\alpha)s_0]}{s_0\sin\alpha}\Big\}\!=\!0,
%\eeq
\beq\label{s0}
F_{s_0}(\alpha)-\cosh(\pi s_0)+\eta(1)\big[1+F_{s_0}(\pi-\alpha)\big]=0,
\eeq
where
\beq\label{F}
F_{s_0}(\xi)\equiv\frac{\sinh(s_0\xi)}{s_0\sin\xi},
\eeq
\beq
\alpha\equiv\left\{\begin{array}{ll}\arccos\frac{u-1}{u+1},~~~&\text{Scenarios 1, 3, 5, 6, and 7},\\
\arccos\frac{u-\cos\theta}{u+1},&\text{Scenarios 2 and 4},
\end{array}\right.
\eeq
\beq\label{eta1}
\eta(1)\equiv\left\{\begin{array}{ll}
0,~~~~&\text{Scenarios 1-5},\\
1,&\text{Scenario 6},\\
-1,&\text{Scenario 7},
\end{array}\right.
\eeq
and $u\equiv m_A/m_B$ is the mass ratio.

In Scenarios 1 and 2, the bound pair, or dimer, is pinned near the vertex (although the dimer's size can be arbitrarily large;
see above). In Scenarios 3 and 4, the dimer is localized near the edge but can move along the edge freely.
In Scenarios 5-7, the dimer is localized near the mirror but can move freely in any direction parallel to the mirror.
The emergence of these shallow bound states will thus be called
\emph{vertex effects}, \emph{edge effects}, and \emph{mirror effects}, respectively.

The vertex effects in Scenarios 1 and 2 will be called \emph{one-leg vertex effect} and \emph{two-leg vertex effect}, respectively.

The edge effects in Scenarios 3 and 4 will be called \emph{one-sheet edge effect} and \emph{two-sheet edge effect}, respectively.

The mirror effects in Scenarios 5, 6, and 7 will be called \emph{Type-I mirror effect}, \emph{Type-IIN mirror effect} and \emph{Type-IID mirror effect},
respectively.
For two resonantly interacting particles near a hard wall with the usual Dirichlet boundary condition,
there are no infinite sequence of bound states
\cite{bouncing2010}. The mirror effects predicted in this Letter differ
in that at least one particle is subject to the \emph{Neumann} boundary condition instead
[see \Eq{mirror Neumann} below].

%In all the seven scenarios, we require the potential at the vertex, edge, or mirror surface
%to have a critical depth such that the ground state wave function of an \emph{isolated} particle A approaches a finite constant
%away from the vertex, edge, or mirror surface.

\begin{figure}
\includegraphics[scale=.67]{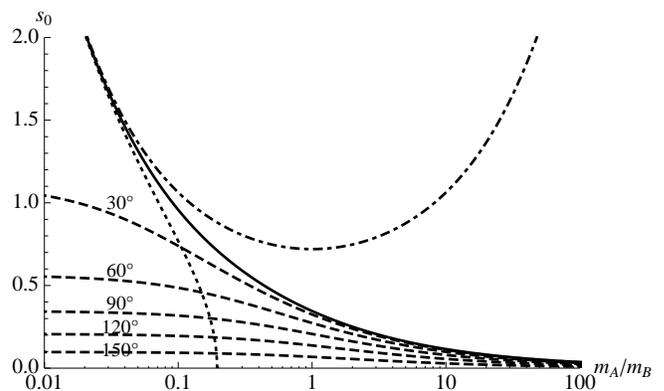}
\caption{\label{fig:s0}
$s_0$ versus the mass ratio.
The sizes of two adjacent shallow bound states differ by a factor of $\lambda=e^{\pi/s_0}$.
Solid line applies to the one-leg vertex effect, one-sheet edge effect, and Type-I mirror effect,
as well as the $\theta\to0$ limits of the two-leg vertex effect and two-sheet edge effect.
Dashed lines: two-leg vertex effect and two-sheet edge effect at various values of the angle $\theta$,
whose value is labeled on each curve.
Dot-dashed line: Type-IIN mirror effect.
Dotted line: Type-IID mirror effect (existing at $m_A/m_B<0.195$ only).
}
\end{figure}

The values of $s_0$ for all the seven scenarios are plotted in Fig.~\ref{fig:s0}.
Note that in Scenario 7, $s_0$ vanishes at $u=u_c\approx0.195$ and there is no real solution for $s_0$
at $m_A/m_B>u_c$, so the Type-IID mirror effect is limited to $m_A/m_B<u_c$ only.
In all the other six scenarios, the infinite sequence of shallow bound states exist for all mass ratios.

We only consider those two-body bound states whose sizes far exceed the range of the interaction;
we treat the range as zero and replace the interaction by a 
Bethe-Peierls (BP) boundary condition [see Eqs.~\eq{1-leg BP}, \eq{1-sheet BP} and \eq{mirror BP} below].
Equation~\eq{s0} ensures that the BP condition is satisfied. The two-body binding energies are given by
\beq\label{Ebinding}
E=-\hbar^2\kappa_n^2/2m_B,
\eeq
where $\kappa_n\equiv\kappa_0\lambda^{-n}$ ($n=0, 1, 2, \dots$).
The parameter $1/\kappa_0$ is determined by the masses of both particles, the microscopic details of the interaction, and the external potential
responsible for the 1D or 2D confinement or the mirror.
It plays the same role as the length parameter in the three-body Efimov effect \cite{Efimov1970,Efimov1971}.

In the following paragraphs we list the equations satisfied by the low energy states
at $|a_\text{eff}|\to\infty$ (for Scenarios 1-4) or $|a|\to\infty$ (for Scenarios 5-7),
and show their bound-state solutions.

In Scenario 1, the two-particle wave function is $\psi=\psi(r_A,\vect r_B)$, where $r_A$ is the distance of particle A from the vertex, and $\vect r_B$ is the position
vector of particle B relative to the vertex. %If the potential for particle A at the vertex has a critical depth
Since the ground state wave function of an \emph{isolated} particle A is a constant on the ray away from the vertex,
$\psi$ satisfies the Neumann boundary condition at low energies
(we assume particle A is tightly confined in a quasi-1D geometry, 
and the two-body energy $E$ relative to the scattering threshold should be much smaller than
the microscopic energy scales such as $\hbar^2/2m_A\epsilon^2$, where $2\pi\hbar$ is Planck's constant,
and $\epsilon$ is the amplitude of zero-point motion of particle A in the transverse directions):
\beq
\lim_{r_A\to0}\frac{\partial\psi}{\partial r_A}=0~~~\text{when }\vect r_B\ne0.
\eeq
$\psi$ also satisfies the Schr\"{o}dinger equation
\beq\label{1-leg Schrodinger}
\Big(-\frac{\hbar^2}{2m_A}\frac{\partial^2}{\partial r_A^2}-\frac{\hbar^2}{2m_B}\nabla_B^2\Big)\psi=E\psi,~~~~\mathcal D>0
\eeq
and the BP boundary condition for the 1D-3D mixture with $|a_\text{eff}|\to\infty$ \cite{MixedD2008}:
\beq\label{1-leg BP}
\psi\propto1/\mathcal D+O(\mathcal D),~~~~\mathcal D\to0,
\eeq
where $\mathcal{D}\equiv\big[|\vect r_B-\vect e(\vect e\cdot\vect r_B)|^2+\frac{u}{u+1}\big(r_A-\vect e\cdot\vect r_B\big)^2\big]^{1/2}$
is a measure of the separation between the two particles \cite{MixedD2008}, and $\vect e$ is the unit vector parallel to the 1D ray along which A is confined.
We find a sequence of bound states (labeled by the integer $n$) with wave functions
\beq
\psi=\sum_{\sigma=\pm1}F_{s_0}\Big(\arccos\frac{\sigma ur_A-\vect e\cdot\vect r_B}{\sqrt{u+1}\,R}\Big)G_{s_0}(\kappa_nR),
\eeq
where $F_{s_0}(\xi)$ is defined in \Eq{F},
$G_{s_0}(y)\equiv y^{-1}K_{is_0}(y)$,
$K_{is_0}(y)$ is the modified Bessel function of the second kind which decays exponentially at large $y$, and
$R\equiv(ur_A^2+r_B^2)^{1/2}$. Their energies are given by \Eq{Ebinding}.

In Scenario 2, let $\psi_\mu(r_A,\vect r_B)$ be the probability amplitude of finding particle A on the $\mu$th arm ($\mu=1$ or $2$)
of the bent line, at distance $r_A>0$ from the vertex, and particle B at position $\vect r_B$ relative to the vertex.
%If the depth of the external potential for particle A at the vertex takes a critical value,
Since the ground state wave function of an \emph{isolated} particle A is a nonzero constant along the two arms,
at low energies $\psi_\mu(r_A,\vect r_B)$ is a continuous and smooth
function of the position of particle A on the bent line:
\beq
\lim_{r_A\to0}\psi_1=\lim_{r_A\to0}\psi_2,~~~
\lim_{r_A\to0}\frac{\partial\psi_1}{\partial r_A}=-\lim_{r_A\to0}\frac{\partial\psi_2}{\partial r_A}.
\eeq
The Schr\"{o}dinger equation and the BP boundary condition are given by Eqs.~\eq{1-leg Schrodinger} and \eq{1-leg BP},
with $\psi$ replaced by $\psi_\mu$ and
$\mathcal D=\big[|\vect r_B-\vect e_\mu(\vect e_\mu\cdot\vect r_B)|^2+\frac{u}{u+1}\big(r_A-\vect e_\mu\cdot\vect r_B\big)^2\big]^{1/2}$ instead.
Here $\vect e_\mu$ is the unit vector parallel to the $\mu$th arm, satisfying $\vect e_1\cdot\vect e_2=\cos\theta$.
We find a sequence of bound states with wave functions
\beq
\psi_\mu=\sum_{\nu=1}^2F_{s_0}\Big(\arccos\frac{(-1)^{\delta_{\mu\nu}}ur_A-\vect e_\nu\cdot\vect r_B}{\sqrt{u+1}\,R}\Big)G_{s_0}(\kappa_nR),
\eeq
where $\delta_{\mu\nu}$ is the Kronecker delta, and $R\equiv(ur_A^2+r_B^2)^{1/2}$. Their energies take the form of \Eq{Ebinding}.

In Scenario 3, we set up a Cartesian coordinate system whose $z$ axis coincides with the edge
of the half-infinite sheet, and whose positive $x$ axis is on the sheet.
The wave function is $\psi(x_A,z_A,x_B,y_B,z_B)$, where $(x_A,0,z_A)$ and $(x_B,y_B,z_B)$ are the coordinates of A and B,
respectively. $\psi$ satisfies the Neumann boundary condition
\beq
\lim_{x_A\to0}\frac{\partial\psi}{\partial x_A}=0~\text{ when }x_B^2+y_B^2+(z_A-z_B)^2\ne0,
\eeq
the Schr\"{o}dinger equation
\beq
\Big[-\frac{\hbar^2}{2m_A}\Big(\frac{\partial^2}{\partial x_A^2}+\frac{\partial^2}{\partial z_A^2}\Big)
-\frac{\hbar^2}{2m_B}\nabla_B^2\Big]\psi=E\psi,~~~\mathcal D>0
\eeq
and the BP boundary condition for the 2D-3D mixture with $|a_\text{eff}|\to\infty$ \cite{MixedD2008}:
\beq\label{1-sheet BP}
\psi\propto\mathcal D^{-1}+O(\mathcal D),~~~\mathcal D\to0,
\eeq
where
$\mathcal D\equiv\big\{y_B^2+\frac{u}{u+1}[(x_A-x_B)^2+(z_A-z_B)^2]\}^{1/2}$
is a measure of the separation between the two particles \cite{MixedD2008}.
We find a sequence of shallow bound states with wave functions and energies
\beq
\psi=e^{ik_zz_c}\sum_{\sigma=\pm1}F_{s_0}\!\Big(\!\arccos\!\frac{\sigma ux_A-x_B}{\sqrt{u+1}\,R}\Big)G_{s_0}(\kappa_nR),
\eeq
\beq\label{1-sheet E}
E=-\hbar^2\kappa_n^2/2m_B+\hbar^2k_z^2/2M,
\eeq
where
$
z_c\equiv(uz_A+z_B)/(u+1)
$
is the $z$ coordinate of the center of mass (C.O.M.),
$M\equiv m_A+m_B$ is the total mass,
$\hbar k_z$ is the $z$ component of the two particles' total linear momentum, and
$R=\big[ux_A^2+x_B^2+y_B^2+\tfrac{u}{u+1}(z_A-z_B)^2\big]^{1/2}.$
The C.O.M. motion in the $z$ direction is decoupled from the remaining degrees of freedom.

In Scenario 4, we set up a Cartesian coordinate system with the edge as the $z$ axis.
The two-body wave function is $\psi_\mu(\rho_A,z_A,\vect r_{B\perp},z_B)$, representing the probability amplitude
of finding particle A on the $\mu$th ($\mu=1, 2$) sheet at distance $\rho_A$ from the edge, with $z$ coordinate $z_A$,
and particle B at position $(\vect r_{B\perp}+z_B\vect e_z)$.
Here $\vect e_z$ is the unit vector along the $z$ axis, and $\vect r_{B\perp}$ is perpendicular to $\vect e_z$.
Let $\vect e_\mu$ be the unit vector parallel to the $\mu$th sheet and perpendicular to the edge: $\vect e_\mu\cdot\vect e_z=0$.
We have again $\vect e_1\cdot\vect e_2=\cos\theta$. The wave function satisfies the continuity and smoothness conditions across the edge
\beq
\lim_{\rho_A\to0}\psi_1=\lim_{\rho_A\to0}\psi_2,~~~\lim_{\rho_A\to0}\frac{\partial\psi_1}{\partial\rho_A}=-\lim_{\rho_A\to0}\frac{\partial\psi_2}{\partial\rho_A},
\eeq
the Schr\"{o}dinger equation
\beq
\Big[-\frac{\hbar^2}{2m_A}\Big(\frac{\partial^2}{\partial\rho_A^2}+\frac{\partial^2}{\partial z_A^2}\Big)
-\frac{\hbar^2}{2m_B}\nabla_B^2\Big]\psi_\mu=E\psi_\mu,~~\mathcal D>0
\eeq
and the BP boundary condition \Eq{1-sheet BP}, but now
\begin{align*}
\mathcal D&=\Big\{\big|\vect r_{B\perp}-\vect e_\mu(\vect e_\mu\cdot\vect r_{B\perp})\big|^2\\
&\quad\quad+\tfrac{u}{u+1}\big[(\rho_A-\vect e_\mu\cdot\vect r_{B\perp})^2+(z_A-z_B)^2\big]\Big\}^{1/2}.
\end{align*}
We find a sequence of bound states with wave functions
\begin{align}
\psi_\mu&=e^{ik_zz_c}\sum_{\nu=1}^2F_{s_0}\Big(\arccos\frac{(-1)^{\delta_{\mu\nu}}u\rho_A-\vect e_\nu\cdot\vect r_{B\perp}}{\sqrt{u+1}\,R}\Big)\nn\\
&\quad\times G_{s_0}(\kappa_nR),
\end{align}
where $R\equiv\big[u\rho_A^2+r_{B\perp}^2+\frac{u}{u+1}(z_A-z_B)^2\big]^{1/2}$
and $z_c\equiv(uz_A+z_B)/(u+1)$.
The energies of these bound states again take the form of \Eq{1-sheet E}.

In Scenarios 5, 6, and 7, we set up a Cartesian coordinate system whose $yz$ plane coincides with the mirror surface,
and whose positive $x$ axis is on the same side of the mirror as particle A.
The two-body wave function is $\psi(x_A,y_A,z_A,x_B,y_B,z_B)$, where $(x_A,y_A,z_A)$ and $(x_B,y_B,z_B)$ are
the coordinates of particles A and B, respectively. In all three Scenarios, $x_A>0$. In Scenarios 6 and 7 we also have $x_B>0$
(but in Scenario 5, $x_B$ may take any value).
The wave function satisfies the Neumann boundary condition
\beq\label{mirror Neumann}
\lim_{x_A\to0}\frac{\partial\psi}{\partial x_A}=0~~\text{when }x_B^2+(y_B-y_A)^2+(z_B-z_A)^2\ne0,
\eeq
the Schr\"{o}dinger equation
\beq\label{mirror Schrodinger}
\Big(-\frac{\hbar^2}{2m_A}\nabla_A^2-\frac{\hbar^2}{2m_B}\nabla_B^2\Big)\psi=E\psi,~~~r>0
\eeq
and the BP boundary condition with $|a|\to\infty$:
\beq\label{mirror BP}
\psi\propto r^{-1}+O(r),~~~r\to0,
\eeq
where $r\equiv\big[(x_A-x_B)^2+(y_A-y_B)^2+(z_A-z_B)^2\big]^{1/2}$ is the distance between the two particles.
In Scenario 6 the wave function must also satisfy the Neumann boundary condition
$
\lim_{x_B\to0}\frac{\partial\psi}{\partial x_B}=0
$
due to a fine-tuning of the mirror potential for particle B at $x_B=0$, whereas in Scenario 7
it should satisfy the Dirichlet boundary condition
$
\lim_{x_B\to0}\psi=0
$
due to a hard wall potential barrier at $x_B=0$.
We find a sequence of bound states with wave functions and energies \cite{symmetry}
\begin{align}
\psi&=e^{ik_yy_c+ik_zz_c}\sum_{\sigma_A=\pm1}\sum_{\sigma_B=\pm1}\eta(\sigma_B)\nn\\
&\quad\times F_{s_0}\Big(\arccos\frac{\sigma_A ux_A+\sigma_B x_B}{\sqrt{u+1}\,R}\Big)G_{s_0}(\kappa_nR),
\end{align}
\beq
E=-\hbar^2\kappa_n^2/2m_B+\hbar^2(k_y^2+k_z^2)/2M,
\eeq
where $y_c\equiv (uy_A+y_B)/(u+1)$, $z_c\equiv(uz_A+z_B)/(u+1)$, 
$R\equiv\big\{ux_A^2+x_B^2+\frac{u}{u+1}\big[(y_A-y_B)^2+(z_A-z_B)^2\big]\big\}^{1/2}$,
$\eta(-1)\equiv1$, and $\eta(1)$ is given by \Eq{eta1}.
%The C.O.M. motion in the $y$ and $z$ directions is decoupled from the remaining degrees of freedom.
If particle A is subject to the Neumann boundary condition \eq{mirror Neumann},
but particle B is subject to a \emph{mixed} condition
$\psi\propto1-x_B/a_{1D}^B$ at $x_B\to0^+$,
the two-body spectrum will be type-IID-like for pair sizes $\gg |a_{1D}^B|$,
type-IIN-like for pair sizes $\ll |a_{1D}^B|$, and exhibits a smooth crossover
in between, analogous to the ``Bose-Fermi crossover" in the three-body Efimov spectrum in a 1D-3D mixture \cite{MixedD20110413}.

One may realize the scenarios discussed above with ultracold atoms, for which one can
use a Feshbach resonance to achieve the two-body BP boundary condition.
For Scenarios 1-5, one may apply species-selective optical dipole potentials
to constrain the motion of atom A but not atom B \cite{MixedD20100126}.
For Scenarios 1-3, the potentials could be produced by some light sources' real images (at the focal plane of a lens or parabolic reflector),
superimposed by a 1D optical lattice.
To realize the Type-I mirror effect (Scenario 5), one may illuminate
the 3D region $x>0$ with a laser (red-detuned for atom A) whose intensity is deliberately enhanced in a layer of thickness $\sim d$ near $x=0$,
such that atom A's ground state wave function approaches a finite constant at $x\gg d$.
To realize the Type-II mirror effects (including Scenarios 6 and 7 but excluding Scenario 5),
one may apply a Double Evanescent Wave mirror \cite{DEW2002} potential $V(x)$
that is strongly repulsive at $x<\epsilon_1$, attractive at $\epsilon_1<x\lesssim\epsilon_2$, and negligible at $x\gg\epsilon_2$. At a critical attraction,
the ground state wave function of a single atom A approaches a finite constant at $x\gg\epsilon_2$.
Then the two-body wave function satisfies \Eq{mirror Neumann} at low energies $|E|\ll\hbar^2/2m_A\epsilon_2^2$.

The Type-IIN mirror effect may be realized with two identical bosons, for which $s_0=0.7202$ and
the scaling factor $\lambda=78.4$ [see Eqs.~\eq{lambda} and \eq{s0}].
This should be contrasted with the scaling factor $22.7$ for the Efimov effect of three identical bosons \cite{Efimov1970, Efimov1971}.
(All other two-body effects predicted above require two distinguishable particles.)

To have a denser two-body spectrum, we need a smaller scaling factor $\lambda$.
This requires a smaller mass ratio $m_A/m_B$ (except for the type-IIN mirror effect where
$\lambda$ is invariant under the interchange of $m_A$ and $m_B$; see Fig.~\ref{fig:s0}).
If one chooses $^6$Li and $^{133}$Cs for A and B respectively, then in the 1-leg vertex effect,
1-sheet edge effect, and type-I mirror effect $\lambda=9.8$,
in the type-IIN mirror effect $\lambda=9.1$, and in the type-IID mirror effect $\lambda=10.7$.

For cold atoms,
the shallow two-body bound states predicted in this Letter
will be free from three-body recombination (if isolated from other atoms and molecules),
and can potentially be \emph{much longer lived} than the Efimov trimers created experimentally
\cite{EfimovExperiment2006, EfimovExperiment20080603, EfimovExperiment20080722,
EfimovExperiment20081017, EfimovExperiment20090120, EfimovExperiment20090127, EfimovExperiment20090625, EfimovExperiment20091005,
EfimovExperiment20100325, EfimovExperiment20101010}.

%One could also use radio frequency pulses to induce transitions between different
%bound levels, which could be useful for information processing.
The spatial mobility of the dimers in Scenarios 3-7
may allow them to be transported and/or manipulated easily.

In summary, we have shown that by tuning two particles near a scattering resonance
and delicately constraining the spatial motion of at least one of them,
one can create many universal giant two-body bound states.
Our scenarios illustrate a close interplay between spatial \emph{geometry} and universal few-body states.
For instance, in Scenarios 2 and 4, by merely increasing the \emph{angle} between the
two rays or sheets to $180^\circ$, one can eliminate the discrete sequence of two-body bound states.

The present work can be extended to large but finite two-body (effective) scattering lengths, or to
situations where the external potential for particle A at the vertex, edge, or mirror is slightly detuned
such that the particle is subject to a large but finite 1D scattering length in 1, 2, or 3 dimensions.
We expect that by reducing the (effective or 1D) scattering length by each factor of $\lambda$,
one shallow two-body bound state disappears.
This is analogous to the three-body Efimov effect (see, eg, Ref.~\cite{Braaten2007}).

By changing the external potentials for particles,
one may realize many more types of universal two-body or few-body effects.
Here we list just a few examples.
\textbf{1.} By confining one particle along a thin circle and allowing another
to move in 3D, one can produce exotic shallow two-body bound states that are entirely determined by macroscopic
parameters such as the mass ratio, the radius of the circle, and the effective scattering length
(there is \emph{no} discrete scaling symmetry in this case).
\textbf{2.} By confining one particle along a hyperbola (a rounded 2-leg vertex) and allowing another to move in 3D,
one can produce shallow bound states with discrete scaling symmetry in the infrared limit,
with a length parameter $1/\kappa_0$ uniquely determined by the mass ratio and the parameters of the hyperbola.
\textbf{3.} One can create new types of vertex effects by confining one particle along the arms of an $n$-leg vertex ($n\ge3$),
which offer many more knobs - the angles between the arms - for controlling the two-body bound states without breaking the discrete scaling symmetry.
\textbf{4.} In the 1-leg vertex effect, one can drive the excitation or de-excitation of a shallow two-body bound state 
by placing another 1D ray nearby and injecting a third particle along it.
Such universal two-body and few-body effects will be studied in the future.
%It appears that we are heading into the age in which the wave functions of systems of two or more particles
%can be \emph{tailored and manipulated} in many ways.

The author thanks Rudi Grimm,  Dean Lee, Yusuke Nishida, and Ran Qi  for discussions.
This work is supported by the NSF Grant PHY-1068511
and by the Alfred P. Sloan Foundation.

\bibliography{TwoBody2012-07}

\begin{thebibliography}{35}
\bibitem[{sym()}]{symmetry}
\bibinfo{note}{In Scenario 6, if the two particles are point-like, be subject
  to different external potentials, and interact with a symmetric two-body
  potential, then the binding energies $E$ are symmetric under the
  \emph{simultaneous} exchange of the masses and the two external potentials,
  but $\kappa_0$ is not, according to \Eq{Ebinding}. Instead,
  $\kappa_0/\sqrt{m_B}$ is symmetric.}
\end{thebibliography}

\end{document}